# Theoretical study and finite element simulation of ZnO/GaAs higher-order Lamb waves for microsensor application in liquid media

Muhammad Hamidullah, Céline Élie-Caille and Thérèse Leblois
FEMTO-ST Institute - Besançon, France

*Abstract*— Lamb waves with dominantly longitudinal displacement component have been reported for sensor application in liquid media. However, they are still limited for the fundamental symmetry mode with low plate thickness-to-wavelength (h/λ) ratio, resulting in large wavelength and low resonance frequency. Here we show the finite element simulation study of higher-order Lamb waves on GaAs and ZnO/GaAs plates. In-plane polarised quasi-longitudinal higher-order Lamb waves are obtained at higher h/λ ratio resulting in higher resonance frequencies. The solid-liquid contact simulation shows the confinement of acoustic energy inside the plate. The result demonstrates the suitability of ZnO/GaAs higher-order quasi-longitudinal modes for microsensor in liquid media.

I. INTRODUCTION

While the applications of the acoustic waves-based sensors have been reported for several decades [1], the recent trend of bio and chemical sensor integration within a microfluidic and lab-on-chip system has risen a new interest of this research field [2]. One particular interest is on plate acoustic waves (PAW) that offer several advantages, such as faster sound velocity and the two free plate surfaces for the separation between the electrical transducer surface and the sensing surface [3]. The bulk acoustic wave (BAW)-based sensor, such as quartz crystal microbalance (QCM) [4] and lateral field excitation thickness-shear waves (LFE-TSW) [5] offer similar advantages; however, the thickness of the plate dictates the resonance frequency of the devices (plate thickness = half of the wavelength). The wavelength in PAW corresponds to the interdigitated transducers (IDTs) geometry so that the resonance frequency can be increased by scaling down the IDTs.

Shear Horizontal Acoustic Plate Mode (SH-APM) is the most common type of PAW for biosensor application in liquid media [6]. SH-APM has in-plane shear horizontal polarisation resulting in the confinement of acoustic energy inside the plate when the surface of the plate is in contact with liquid media. Lamb wave is another type of PAW [7]. In general, Lamb wave is elliptically polarised with two particle displacement components: longitudinal and shear-vertical. Under specific plate thickness to wavelength (h/λ) ratio, Lamb wave has a dominantly in-plane longitudinal component with close to zero shear vertical component. For fundamental symmetry Lamb waves ($S_0$), the h/λ ratio is generally less than 0.1; thus the wavelength is ten times larger than the plate thickness. Recently, higher-order Lamb waves with a quasi-longitudinal component in ST-cut Quartz [8] at a higher h/λ ratio, resulting in a reduction in wavelength to achieve higher resonance frequency.

Gallium Arsenide (GaAs) is a unique material due to its piezoelectric and optical properties for potential application of acoustic [5] and photonic sensor [9] integrated into the same substrate. GaAs biosensor can be integrated with a microfluidic channel by direct bonding process [10]. However, the piezoelectricity of GaAs is relatively smaller compared to those commonly used piezoelectric materials. Deposition of thin-film with higher piezoelectricity such as ZnO can increase the electromechanical coupling coefficient $K^2$ of the generated acoustic waves [11].

In this paper, quasi-longitudinal higher-order Lamb waves based on GaAs and ZnO/GaAs will be theoretically studied by finite element simulation. The effect of ZnO deposition to $K^2$ and displacement field profile will be evaluated. Finally, the solid-liquid contact of GaAs/ZnO Lamb waves will be simulated to verify the suitability for sensor application in liquid media.

## II. GaAs Lamb waves

COMSOL Multiphysics-eigenfrequency studies were used to obtain the dispersion curves of the first six Lamb waves of {100}-cut GaAs propagating along the ⟨110⟩ direction. There are two criteria for the Lamb waves with quasi-longitudinal characteristic: the wave velocity is closed to bulk longitudinal velocity, and the mode is lowly dispersive [8]. Based on those criteria, first and second symmetry higher order modes (QL-$S_1$ and QL-$S_2$) are identified as quasi-longitudinal modes at $h/\lambda$ ratio of 0.6 and 1.2, respectively, as shown in figure 1.

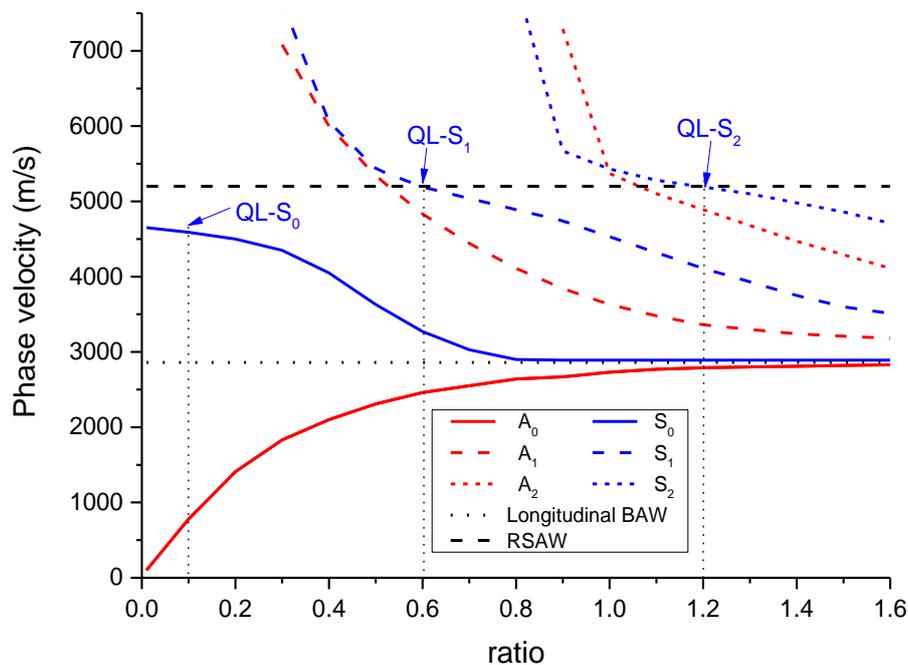

Fig. 1. Lamb wave dispersion curve of {100}-cut GaAs propagating along the ⟨110⟩ direction.

The plate depth longitudinal ($u_1$) and shear vertical ($u_3$) displacement component profile of the fundamental symmetry Lamb wave QL-$S_0$, QL-$S_1$ and QL-$S_2$ are shown in figure 2. As shown in figure 2, all three modes have dominantly $u_1$ component at the surfaces, however, the $u_3$ component for QL-$S_0$ mode is comparatively more significant than QL-$S_1$ and QL-$S_2$. Higher $u_3$ component on the surface of the plate will increase the acoustic energy conversion into pressure wave in the liquid, resulting in higher damping loss. For QL-$S_1$ and QL-$S_2$, $u_3$ components are close to zero so that the acoustic energy loss into the liquid will be minimised.

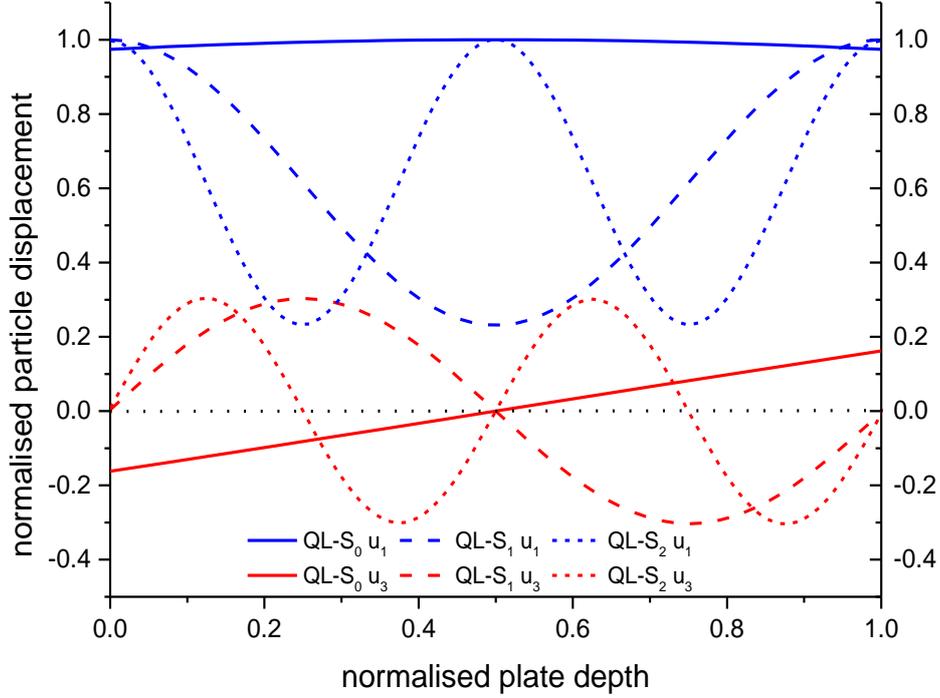

Fig. 2. Displacement depth profile of QL-$S_0$, QL-$S_1$ and QL-$S_2$

The $K^2$ of the three different modes are summarised in table 1 for two different coupling configuration. The deposition of an additional metal layer on the surface opposite of the plate surface with IDTs will result in different electrical boundary condition. The metal-substrate-transducer (MST) and substrate-transducer (ST) are coupling configuration with and without additional metal layer. For simplification, floating electrical boundary conditions are used to represent a massless metal layer in the simulations. However, in the real application, finite thickness metal layer will be used which will give not only an electrical boundary condition but also add to the total thickness of the plate.

TABLE I.   $K^2$ OF GaAs LAMB WAVES

| Modes | GaAs thickness (h) | Wavelength ($\lambda$) | Ratio h/$\lambda$ | Operating Frequency | $K^2$ ST | $K^2$ MST |
|---|---|---|---|---|---|---|
| QL-$S_0$ | 30 μm | 300 μm | 0.1 | 15.8 MHz | 0.025% | 0.151% |
| QL-$S_1$ | 30 μm | 50 μm | 0.6 | 105.5 MHz | 0.057% | 0.057% |
| QL-$S_2$ | 30 μm | 25 μm | 1.2 | 207.8 MHz | 0.018% | 0.028% |

As shown in table 1, the $K^2$ for all three modes with ST configuration are comparable. The $K^2$ of QL-$S_0$ mode is six times higher for MST configuration. However, there is no significant increase for QL-$S_1$ and QL-$S_2$ modes with MST configuration. In [7], thin molybdenum (Mo) metal layer is used as bottom electrode on AlN thin film Lamb waves devices (Mo/AlN/IDT coupling configuration). The result shows improvement in the excitation efficiency of $S_0$ Lamb waves. The report is consistent with the improvement in $K^2$ of the simulated MST coupling configuration for the QL-$S_0$ mode.

III.   LAYERED THIN FILM ZnO/GaAs SUBSTRATE

Hexagonal piezoelectric thin film such as AlN and ZnO has been reported as a promising method to excite acoustic waves on non-piezoelectric material such as silicon [12] and to increase the $K^2$ on material with low piezoelectric constant [13]. In a thin film-substrate layered structure, there are four possible coupling configurations as shown in figure 3. $K^2$ for four different configurations are calculated for 1 μm ZnO thin film deposited on 30 μm GaAs, and the results are summarised in table 2. In general, the deposition of ZnO on GaAs increase the $K^2$ of all modes, except for the STF configuration. An improvement of more than one order in $K^2$ is achieved by the SMFT coupling configuration.

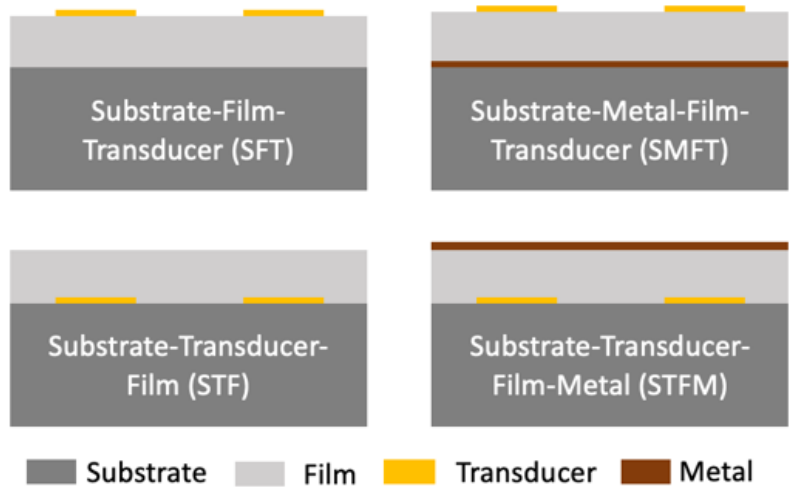

Fig. 3. Layered thin film ZnO/GaAs substrate configurations

TABLE II. $K^2$ OF ZNO/GAAS LAMB WAVES

| Modes | Operating Frequency | $K^2$ SFT | $K^2$ STF | $K^2$ SMFT | $K^2$ STFM |
|---|---|---|---|---|---|
| QL-$S_0$ | 15.9 MHz | 0.038% | 0.013% | 0.389% | 0.364% |
| QL-$S_1$ | 105.8 MHz | 0.227% | 0.038% | 0.607% | 0.417% |
| QL-$S_2$ | 211.5 MHz | 0.237% | 0.009% | 0.579% | 0.351% |

While the deposition of ZnO improves the $K^2$, it may also alter the displacement field profile where the $u_3$ displacement component is no longer close to zero on the surface of the plate. Moreover, a non-homogenous ZnO/GaAs will result in unsymmetrical displacement profile. The plate displacement profile of ZnO/GaAs plate is shown in figure 4. As shown in figure 4, the $u_1$ component is still dominant for all modes, thus all three modes are still dominantly longitudinal modes even with the addition of ZnO thin film.

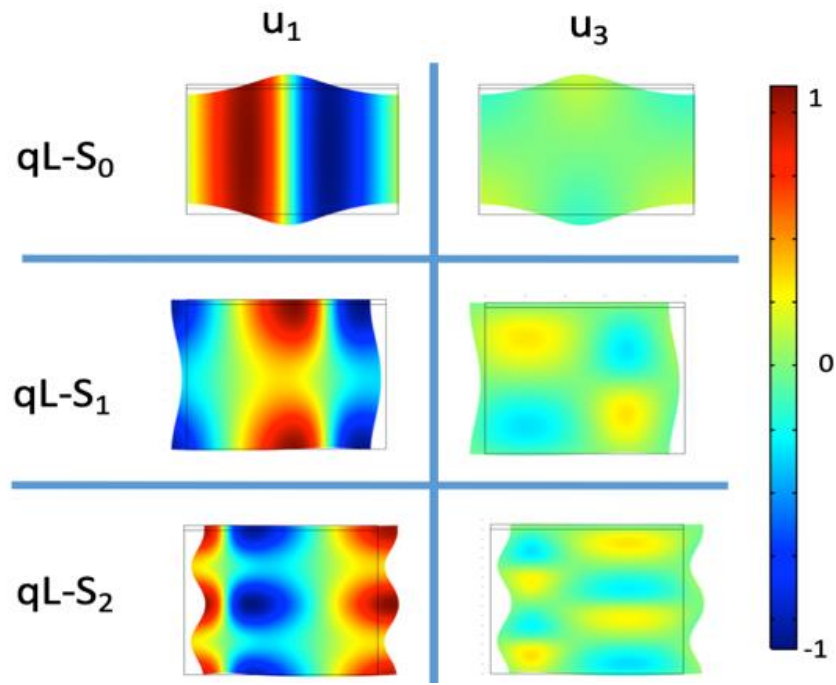

Fig. 4. Normalised depth $u_1$ and $u_3$ displacement profiles of ZnO/GaAs plate for the SMFT configuration. Images are not in scale. All three plates have the same thickness of 31 μm but different wavelength. The width of the plates are equal to one wavelength.

While the three modes are quasi-longitudinal modes, however, it is important to evaluate the effect of ZnO film to the $u_3$ displacement on the plate surfaces. Figure 5 shows the $u_3$ absolute displacement profile normalised to the maximum $u_1$ displacement, with focus on the plate surfaces. As shown in figure 5, the surface displacement profile is no longer symmetrical where the $u_3$ displacement at the GaAs surface is higher than at the ZnO surface for all three modes. The increase in the $u_3$ displacement on one side of the plate is more prominent in for QL-$S_1$ and QL-$S_2$ modes. However, the displacement is still lower than the QL-$S_0$ mode, thus it is expected that QL-$S_1$ and QL-$S_2$ will have less viscous damping when the plate surface is in contact with liquid.

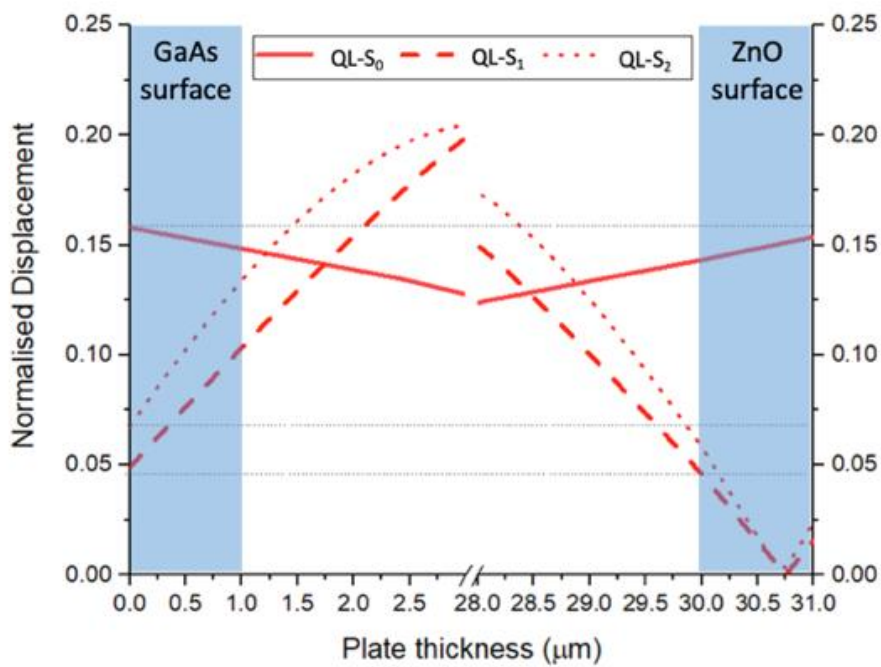

Fig. 5. Absolute $u_3$ displacement on the plate surfaces, normalised to maximum $u_1$ displacement.

## IV. SOLID-LIQUID CONTACT

COMSOL eigenfrequency and frequency domain studies were performed to study the acoustic energy loss when the plate is in contact with liquid media. For the ZnO/GaAs plate, the IDTs is located in the ZnO surface, while the opposite GaAs surface will be in contact with media. The simulation result of ZnO/GaAs plate-liquid contact is shown in figure 5. As shown in figure 6, the acoustic energy is confined in the plate for all modes, however, for the QL-$S_0$, there is a significant conversion of acoustic energy into pressure wave in the liquid media.

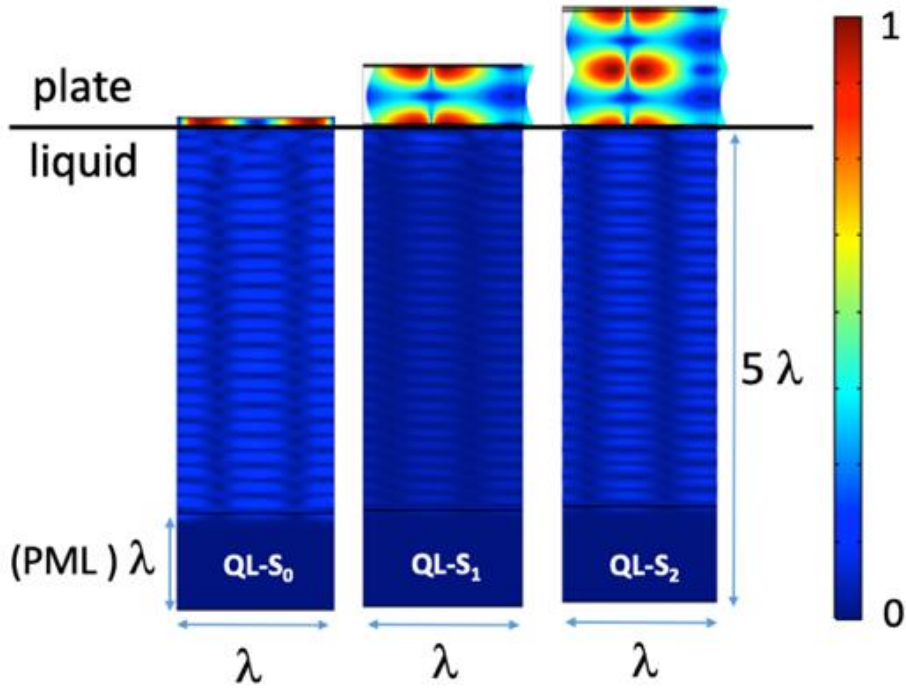

Fig. 7. Plate-liquid contact eigenfrequency simulation results with a perfectly matched layer (PML) at the bottom of the liquid to represent semi-infinite half-space. The scale bar shows the normalised total displacement. Images are not in scale. All three plates have the same thickness of 31 μm but different wavelengths. The difference plate thickness on the images corresponds to the h/λ ratio.

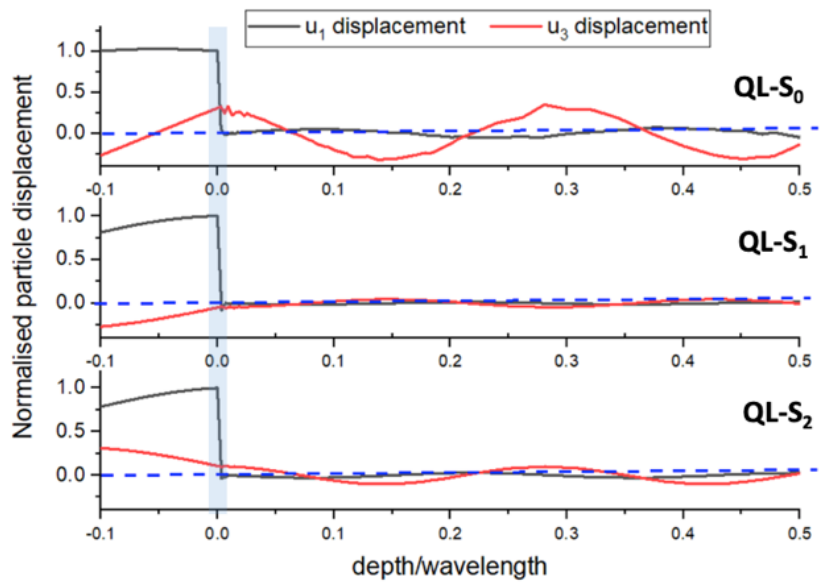

Fig. 8. Propagation and conversion of Lamb waves into pressure wave at the plate-liquid interface (x-axis = 0). The negative and positive x-axis represent the normalised depth/wavelenth of the plate and liquid respectively

The propagation and the conversion of Lamb waves into pressure waves on the plate-liquid contact is shown in figure 7. As shown in figure 7, the $u_1$ particle displacement is confined in the plate and the displacement is reduced exponentially at the plate-liquid interface. This interface layer will act as viscous-mass loading when the plate is in contact with liquid. Furthermore, the $u_3$ displacements are converted into pressure wave in the liquid, resulting in acoustic energy loss. As expected, QL-$S_0$ has higher normalised pressure wave displacement in the liquid due to higher $u_3$ displacement on the surface of the plate.

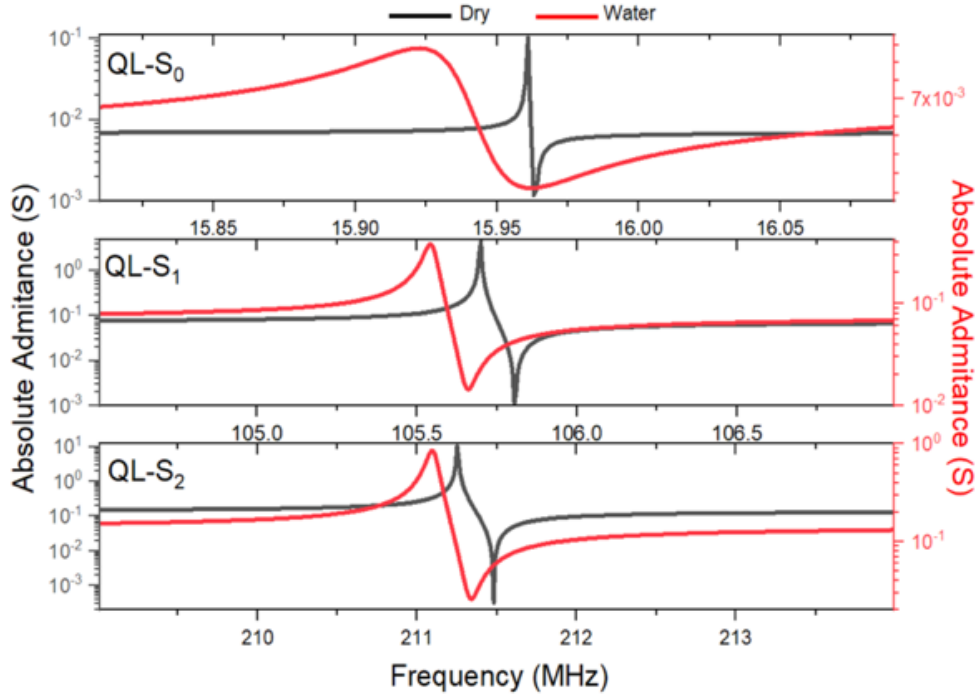

Fig. 9. Frequency domain simulation results for QL Lamb waves resonator

The frequency domain simulation results of a single IDT resonator are shown in figure 8. As shown in figure 8, the resonance frequency is lower when the plate is in contact with the liquid due to viscous-mass loading effect as previously shown and explained in figure 7. Since the QL Lamb waves are not purely in-plane polarised, there are still acoustic energy losses into the liquid due to the $u_3$ displacement component. The acoustic energy loss can be observed by comparing the absolute admittance-frequency curve with and without liquid contact. As previously shown in figure 7, while all modes show acoustic energy losses into liquid media, QL-$S_1$ and QL-$S_2$ have minimum losses and the acoustic resonance signals in figure 8 are distinguishable. Therefore, the study shows that ZnO/GaAs higher order QL-$S_1$ and QL-$S_2$ Lamb waves are feasible for sensing application when one of the plate surfaces is in contact with liquid media.

## V. CONCLUSION

The potential application of ZnO/GaAs for microsensor in liquid media is reported in this paper based on higher order quasi-longitudinal Lamb waves. The COMSOL finite element simulation studies were performed to obtain the dispersion curve and the design parameter in term of plate thickness-to-wavelength ratio. The QL-$S_1$ and QL-$S_2$ GaAs Lamb waves are achieved at h/λ ratio of 0.6 and 1.2, respectively. Furthermore, thin film ZnO/GaAs plate are studied to increase the $K^2$. The improvement of more than one order in $K^2$ is shown for SMFT coupling configuration. Finally, the plate-liquid simulations are performed to verify the confinement of acoustic energy on the plate and the acoustic energy conversion into the pressure wave in liquid media. The eigenfrequency and frequency domain study show that QL-$S_1$ and QL-$S_2$ Lamb waves of ZnO/GaAs plate are suitable for microsensor application in liquid media.

ACKNOWLEDGMENT

This conference proceeding is part of SmOoC project that has received funding from the European Union's Horizon 2020 research and innovation programme under the Marie Skłodowska-Curie grant agreement No. 844135